\documentclass{iucr}           
\journalcode{A}
                   \def\href#1{\relax}\let\foo\caption
\ifPDF
  \RequirePackage{hyperref}
  \PassOptionsToPackage{pdftex,bookmarksopen,bookmarksnumbered}{hyperref}
\fi
\let\caption\foo

     \paperprodcode{a000000}      
     \paperref{xx9999}            
     \papertype{FA}               

     \paperlang{english}          
     \journalyr{2017}
     \journalreceived{\relax}
     \journalaccepted{\relax}
     \journalonline{\relax}

\usepackage{amsmath}
\usepackage{amssymb}
\usepackage{mathtools}
\usepackage{graphicx}
\usepackage{oubraces}
\usepackage[colorinlistoftodos]{todonotes}
\usepackage{amsthm}

\DeclareMathOperator{\Tr}{Tr}
\DeclareMathOperator{\Diag}{Diag}
\DeclarePairedDelimiter\abs{\lvert}{\rvert}%

\begin{document}                  



\title{A Markov theoretic description of stacking disordered aperiodic crystals including ice and opaline silica}
\shorttitle{Aperiodic crystals as Markov chains}


	\cauthor[a]{A. G.}{Hart}{a.hart@bath.ac.uk} \\
     	\author[b]{T. C.}{Hansen} 
     	\author[c]{W. F.}{Kuhs} 
     
        \aff[a]{University of Bath, Bath, UK}
    	\aff[ab]{Institut Laue-Langevin, Grenoble, France}
	    \aff[c]{GZG Abt. Kristallographie, Universit\"{a}t G\"{o}ttingen, Germany}


\shortauthor{Hart, Hansen and Kuhs}



\keyword{Aperiodic}
\keyword{Markov}
\keyword{Chaotic Crystallography}



\maketitle                        

\begin{abstract}
We review the Markov theoretic description of 1D aperiodic crystals, describing the stacking-faulted crystal polytype as a special case of an aperiodic crystal. Under this description we generalise the centrosymmetric unit cell underlying a topologically centrosymmetric crystal to a reversible Markov chain underlying a reversible aperiodic crystal. We show that for the close-packed structure,  almost all stackings are irreversible when the interaction \emph{reichweite} $s > 4$.  Moreover, we present an analytic expression of the scattering cross section of a large class of stacking disordered aperiodic crystals, lacking translational symmetry of their layers, including ice and opaline silica (opal CT). We then relate the observed stackings and their underlying \emph{reichweite} to the physics of various nucleation and growth processes of disordered ice. We proceed by discussing how the derived expressions of scattering cross sections could signiﬁcantly improve implementations of Rietveld's refinement scheme and compare this $Q$-space approach to the pdf-analysis of stacking disordered materials.
\end{abstract}


\section{Introduction}

1D aperiodic crystals are similar to ordinary crystals by virtue of being translationally symmetric in two independent directions yet differ by being aperiodic in the third. Consequently, they cannot be described by just an underlying unit cell and lattice, suggesting the need to go beyond the usual language and formalism of crystallography to describe them.

The contemporary description of these crystals begins with a paper by \citeasnoun{hendricks1942x}, proposing that the aperiodic direction is composed of a sequence of layers where the probability of a layer being a certain type depends on some finite number of near neighbour layers. \citeasnoun{jagodzinski} later assumed this dependence was caused by an interatomic interaction with constant range he called \emph{reichweite}, which he denoted with the integer $s$ if and only if a layer's type depends on $s$ preceding layers. This sequence of layers was described by \citeasnoun{MarkovCrystal} as a sequence of random variables where a given variable depends on only finitely many preceding variables; which the authors recognised as a Markov chain of order $s$. They successfully analysed this description of aperiodic crystals by partitioning the layer chain into blocks comprising $s$ adjacent layers and noting each block depends only on the immediately preceding block, hence reducing the chain to first order.
The approach taken by \citeasnoun{MarkovCrystal} has been used practically by
\citeasnoun{cherepanova2004simulation} to
 model graphite like materials as Markov chains, allowing them to study their X-Ray diffraction pattern. \citeasnoun{hostettler2002structure} expressed the average structure factor of the aperiodic \emph{orange crystal} HgI$_{2}$ using the same principle, hence could study the crystal's diffraction pattern. 
 
 The probability of a particular layer being a certain type depending on some nearest neighbour interaction is strongly analogous to the axial next nearest neighbour Ising (ANNNI) model, as noticed by
\citeasnoun{christy1989short}
and
\citeasnoun{shaw1990nature}. \citeasnoun{christy1989short} went further and used the ANNNI model to compare the suite of polytypes generated for sapphirine and wollastonite each with \emph{reichweite} equal to 4. An overview of aperiodic crystal theory is provided by
\citeasnoun{estevez2007powder}
and the elementary treatment of Markov theory to describe 1D aperiodicity is presented by
\citeasnoun{welberry2004diffuse}. 

In addition to simple crystal structures, aperiodic polytypes have been studied by \citeasnoun{varn2004finite} who simulated transforms between ordered and disordered polytypes, specifically the solid state
transformation of annealed ZnS crystals from a
2H to a 3C. Furthermore, \citeasnoun{varn2013machine} and \citeasnoun{varn2004finite} have made significant progress relating the statistics of stacking faults distributed throughout a general aperiodic crystal to the crystal's scattering pattern using the tools of information theory. The authors cast the problem of finding the simplest possible process that could give rise to an aperiodic crystal exhibiting the observed scattering pattern as defining the crystal's $\varepsilon$-machine, which may be represented as a directed graph or a hidden Markov model (HMM). The HMM describes a Markov process with states hidden from the observer, but each emitting an observable (in this context a layer type) following some probability distribution depending on the state. The resultant sequence of observables (layer types) may not be a Markov chain of any order, and in this case would describe an aperiodic crystal with infinite $reichweite$. However, all Markov chains of finite order can be described by some HMM, so the formulation of an aperiodic crystal as a HMM generalises the Markov theoretic description of aperiodic crystals.  

This work on creating an information theoretic description of aperiodic crystals has culminated in the new field of \emph{chaotic crystallography} summarised elegantly by
\citeasnoun{Varn201547}, who invite the reader to view the work on aperiodic crystals as the foundations of a new generalised crystallography, encompassing the study of disordered materials as well the as translationally symmetric and ordered structures studied by so-called \emph{classical crystallography}. 

Our present paper builds upon chaotic crystallography by first reviewing the description of 1D aperiodic crystals that encompasses both periodic and aperiodic polytypes. Under this description we generalise the notion of a topologically centrosymmetric crystal having an underlying centrosymmetric unit cell to a reversible aperiodic crystal having an underlying reversible Markov chain. By extending the notion of acentricity to aperiodic crystals, it may be possible to explain the physical properties of aperiodic crystals in formal analogy to the properties of acentric periodic crystals. Examples include the piezoelectric effect exhibited by acentric crystals which may have an analogous effect exhibited by aperiodic crystals with an underlying irreversible Markov chain.

This paper also adds to the chaotic crystallographer's mathematical toolbox. We use the matrix formulation of Markov theory to build on the work of \citeasnoun{jagodzinski1954symmetrieeinfluss} to construct new analytic expressions for the differential scattering cross section of 1D aperiodic crystals with finite \emph{reichweite} $s$. 
Several similar derivations do of course exist, including that of \citeasnoun{kakinoki1954intensity} who reached a different expression, that is mathematically interesting but more cumbersome than ours. There is also the expression for the average structure factor product of a single layer derived by \citeasnoun{Allegra}, which is superseded by our expression for the entire scattering cross section. More recently \citeasnoun{riechers2015pairwise} produced a deep theoretical paper relating the scattering pattern of a close-packed structure to its underlying $\varepsilon$-machine, in doing so, produced the most general possible expression for the scattering cross section of a close-packed structure. However, their expression was derived by exploiting a feature of the close-packed structure that does not apply to all aperiodic crystals, namely that the layer types are translations of one another in real space. This assumption is very powerful, as the authors demonstrate by deriving closed form expressions for the diffraction pattern of crystals with independent and identically distributed layers, crystals exhibiting random growth and stacking faults, and crystals exhibiting Shockley-Frank stacking faults. However, the assumption that a crystal's constituent layers differ only by a translation in real space does not hold for the aperiodic ice described by \citeasnoun{Hansen2008}, nor does it hold for aperiodic opal, composed of a disordered sequence of cristobalite and tridymite layers. Motivated by aperiodic ice and opal, we have derived a general expression for the cross section for an aperiodic crystal with finite $s$ that does not assume different layer types have translational symmetry. Since we are interested in finite $s$, we need only consider Markov theory without worrying about the theory of HMMs. However, for completeness, we do include a derivation in Appendix \ref{HMM_CS} of the cross section of the general aperiodic crystal described by a HMM without the assumption of translationally symmetric layer types.

 The computational cost of evaluating our cross section is compared to the cross section developed by \citeasnoun{hendricks1942x} and \citeasnoun{warren1959x} then implemented by
\citeasnoun{treacy1991general} in the software package DIFFaX.  

We argue that our expression is computationally efficient, and could be used to significantly improve several modern uses of \citeasnoun{Rietveld}'s scheme to refine theoretical cross sections of real aperiodic crystals like ice or opal. Examples include \citeasnoun{kuhs2012extent}, \citeasnoun{PhysRevB.34.3586} and \citeasnoun{Hansen2008} who needlessly employed a Monte Carlo simulation to estimate the scattering cross section of ice which could have been estimated more accurately at a much greater speed using our expression for the cross section. 

\section{Forging the Markov chain} 
\label{Forge}

We begin by representing a 1D aperiodic crystal as a sequence of layers, each of a finite set of distinct layer types.
The probability that a layer is a particular type is assumed to depend on the type of each of finitely many consecutive preceding layers. The number of interacting preceding layers is denoted $s$ and called \emph{reichweite} as coined by \citeasnoun{jagodzinski}.

 Following the method employed by \citeasnoun{MarkovCrystal}, the crystal is partitioned into blocks each comprising $s$ consecutive layers - then the set of distinct blocks is indexed with the set $B$. Blocks go by different names in different papers, including \emph{structural motifs} in the work of \citeasnoun{michels2013analyzing}, but we will continue to refer to them as \emph{blocks}. We notice that a block's type depends only on the block immediately preceding it, revealing that a complete sequence of blocks is generated by a first order Markov process. For those interested in forming an analogy with the ANNNI model, consider a nearest neighbour Ising chain with $m$ discrete spin states, rather than just 2 states; up and down. The spin state at each site is analogous to the block type of a particular set of $s$ consecutive layers, hence $m$ is equal to the number of distinct block types $|B|$.
The $m$ state ANNNI model is usually analysed using the \emph{transfer matrix} method \cite{baxter2007exactly}, which is essentially what we proceed with here. Making use of the probabilist's lexicon, we define the \emph{transition matrix} (another name for the transfer matrix) $\boldsymbol{\Xi}$ with elements $\xi_{ij}$ representing the probability that a block is type $j \in B$ given its predecessor is type $i \in B$. 
Next, elementary results of Markov theory reveal that the probability of a block being type $j$ given that the $\nu$th preceding block is type $i$ is the $ij$th element of the matrix $\boldsymbol{\Xi}^{\nu}$; the transition matrix raised to the power $\nu$.

We now seek to compute the probability $\pi_{i}$ that a block sampled from an aperiodic crystal is type $i$. Here, to \emph{sample a crystal} is to randomly, and with equi-probability, select a position from an infinite crystal and observe which type of block is at that position. By assumption, the probability of sampling a $j$ block is related to the probability of the previous block being type $i$, which gives rise to the self-consistency condition
\begin{equation}
\pi_{j} = \sum_{i \in B} \pi_{i}\xi_{ij}
\end{equation}
which may be expressed in matrix notation
\begin{equation}
\boldsymbol{\pi} \boldsymbol{\Xi}= \boldsymbol{\pi}
\end{equation}
where $\pi_{i}$ are elements of the so-called stationary distribution vector $\boldsymbol{\pi}$, a left eigenvector of the transition matrix $\boldsymbol{\Xi}$ with corresponding eigenvalue $1$. The row vector $\boldsymbol{\pi}$ contains the elements of a probability distribution, so satisfies the normalisation condition
\begin{equation}
\sum_{i \in B}\pi_{i} = 1.
\end{equation}

\subsection{Analysis of the block chain}



\subsection{Block pair-correlation function} \label{MatrixVectMult}

The pair-correlation function
$G_{ij}(\nu)$ with $ij \in B$ is defined as the probability that a layer block sampled from a crystal is type $i$ and the block $\nu$ blocks ahead is type $j$. We observe that
$G_{ij}(\nu)$ are the components of the matrix
\begin{equation}
\boldsymbol{G}(\nu) = \Diag(\boldsymbol{\pi})\boldsymbol{\Xi}^{\nu} \label{pipixi}
\end{equation}
where $\Diag(\boldsymbol{\pi})$ is the diagonal matrix with diagonal entries the components of the vector $\boldsymbol{\pi}$.

\subsubsection{Convergence to a stationary distribution} \label{Convergence}

We shall analyse $\boldsymbol{G}(\nu)$ as a sequence in $\nu$ under the technical conditions that the transition matrix $\boldsymbol{\Xi}$ generates a chain that is \emph{positive recurrent}, \emph{irreducible} and \emph{aperiodic}; this allows us to build a description of aperiodic crystals which can be later extended to polytypes.
We justify these three conditions by assuming an aperiodic crystal has three respective physical properties. The first, is that after observing that a block is type $i$, the mean number of blocks after which another is type $i$ is finite. The second, is that after observing that a block is type $i$ there will certainly be a block of type $j$ somewhere ahead of the block $i$. Thirdly, all block types have \emph{period} $k = 1$, where the period $k$ is defined 
\begin{equation}
k = \gcd\{ \nu > 0 : (\boldsymbol{\Xi}^{\nu})_{ii} > 0 \}
\end{equation}
where $\gcd$ stands for greatest common divisor.
It suffices to assume any block type has non-zero probability of following any other to satisfy all three condition. For a chain that is positive recurrent, irreducible and aperiodic, each row of the matrix $\boldsymbol{G}(\nu)$
converges as $\nu \to \infty$ with exponential order to the same unique steady state row vector $\boldsymbol{\pi}$; specifically the convergence is $\mathcal{O}(\abs{\lambda_{2}}^{\nu})$ where $\lambda_{2}$ is the second largest eigenvalue  of the transition matrix $\boldsymbol{\Xi}$. Proof is included in many books on Markov theory like \citeasnoun{MarkovChains}. This result is useful because it allows a crystallographer to estimate how distant a pair of blocks must be before it becomes reasonable to approximate their separation as infinite, and therefore assume their types are uncorrelated. There is however some subtlety to this, which is covered in more detail by \citeasnoun{riechers2015pairwise}. 

\subsubsection{Entropic density of aperiodic crystals}

 \citeasnoun{MarkovCrystal} observed that 1D aperiodic crystals can be represented by a finite order Markov chain, a special type of an $\varepsilon$-machine considered by \citeasnoun{Varn201547} in their discussion of chaotic crystallography. The entropic density $h$ discussed by Varn \& Crutchfield can be expressed in terms of the matrix formalism so far developed 
\begin{equation}
h = -\sum_{ij} \pi_{i}\xi_{ij}\log(\xi_{ij})
\end{equation}
for ease of computation in practical scenarios. The entropy rate can be informally thought of as measuring an aperiodic crystal's \emph{disorder per unit length} in the aperiodic direction, maximising at $h = \log(|B|)$ when any block could follow its predecessor with equi-probability, and minimising at $h = 0$ in the limit as the crystal becomes perfectly periodic. $h$ can be computed in the same way for ensembles that are usually thought of as periodic crystals with stack defects, and provides a measure of how much order there is to the distribution of stack defects. For example, a crystal comprising stack defects that tend to be nearly equally spaced will have a lower entropy rate than one with the same frequency of stack defects distributed with equi-probability throughout the crystal. One might expect from thermodynamic considerations that the entropic density will increase with increasing temperature.

\subsubsection{Aperiodic crystals generalise polytypes}

\citeasnoun{varn2013machine} and \citeasnoun{riechers2015pairwise} have remarked that a polytype comprising a periodic array of unit cells is in fact a special case of an aperiodic crystal. In the formalism used here, a Markov chain with transition matrix $\boldsymbol{\Xi}$ underlies a polytype if and only if its state space can be reduced to a closed communicating class $\mathcal{C}$ with the additional property that $\xi_{ij} \in \{0, 1\} \ \forall ij \in \mathcal{C}$. Less formally, any Markov chain that eventually underlies a sequence of blocks, where the probability of some block following the next is either unity or zero is a polytype.

Now, recall that 1D aperiodic crystals are described by \emph{positive recurrent}, \emph{irreducible} and \emph{aperiodic} chains and note that this cannot be said of the chains that underlie polytypes. However, we can reduce a polytype's underlying chain to the closed communicating class $\mathcal{C}$ which is both \emph{positive recurrent} and \emph{irreducible} forming sufficient conditions for the stationary state vector $\boldsymbol{\pi}$ to exist \cite{MarkovChains}. Conveniently, a Markov chain underlying a polytype differs from that underlying an aperiodic crystal by reducing to a chain that is \emph{periodic}.

\subsection{An Analytic expression for differential scattering cross section of 1D aperiodic crystals}

Neutron and X-ray diffraction experiments can measure a 1D aperiodic crystal's differential scattering cross section, which we seek to express as concisely as possible. We begin by noting that a 1D aperiodic crystal possesses long range order across the basel plane spanned by two primitive lattice vectors $\vec{a}$ and $\vec{b}$ and is aperiodic along the direction plane's normal $\vec{c}$. We let $N_{a}$ and $N_{b}$ denote the number of unit cells across the respective lattice vectors spanning the basal plane, and note that the crystal's cross sectional area is therefore $N_{a} \times N_{b}$. Our notation has been standard so far - but we now break convention and let $N_{c}$ denote the number of blocks (rather than unit cells) stacked in the $\vec{c}$ direction, which together comprise the entire crystal (We should note here that our crystals are now cuboids). Returning to convention, we denote positions in reciprocal space by
\begin{equation}
\vec{Q} = 2\pi(h\vec{a}^* + k\vec{b}^* + l\vec{c}^*)
\end{equation}
where $\vec{a}^*$, $\vec{b}^*$ and $\vec{c}^*$ are primitive lattice vectors and $h,k,l$ are real numbers. 
For a macroscopic crystal where $N_{a}$ and $N_{b}$ are large we have an expression for the differential  cross section developed by \citeasnoun{PhysRevB.34.3586}
\begin{align}
&\frac{d\sigma(\vec{Q})}{d\Omega} = \frac{\sin^2(N_a\pi h)}{\sin^2(\pi h)}\frac{\sin^2(N_b\pi k)}{\sin^2(\pi k)} \times \label{BerlinerExpression} \\
&\sum^{N_{c}}_{m_{3} = -N_{c}}(N_{c} - \abs{m_{3}})Y_{m_3}(\vec{Q})e^{2 \pi i m_{3} l}.
\nonumber 
\end{align}
Here, $Y_{m_{3}}(\vec{Q})$ is the average structure factor product which has expression
\begin{align}
Y_{m_{3}}(\vec{Q}) = \sum_{i \in B }\sum_{j \in B}G_{ij}(m_{3})F_{i}(\vec{Q})F^{*}_{j}(\vec{Q}) \label{SFExpression}
\end{align}
where $G_{ij}(m_{3})$ is exactly the pair-correlation discussed in section \ref{MatrixVectMult}; the probability of finding a layer block of type $j$ separated by a distance $m_{3}c$ from a block of type $i$. Finally, $F_{i}$ is the structure factor of a type $i$ block, $F_{i}^{*}$ is its complex conjugate, and $h_{0}$, $k_{0}$ are nodal lines. Berliner and Werner's 
expression \eqref{BerlinerExpression} was derived using the works of \citeasnoun{wilson1942imperfections}, \citeasnoun{hendricks1942x} and \citeasnoun{jagodzinski}. Further, \eqref{BerlinerExpression} has been used practically to study the cross section of computer generated statistically faulted crystals by \citeasnoun{PhysRevB.34.3586} themselves in an effort to analyse stacking-faults in the 9R lattice and compare the results of simulation to those measured for Li at 20K.
More recently, in a pair of papers by \citeasnoun{Hansen2008} \& (2008$b$), equation \eqref{BerlinerExpression} was used to compute the differential scattering cross section of stack-faulted cubic ice.

Since \eqref{BerlinerExpression} is widely used in simulations, it would be useful to use the matrix formalism of Markov theory to derive an equivalent analytic expression that can be efficiently evaluated by a computer. Starting from equations \eqref{BerlinerExpression} and \eqref{SFExpression} then building on work by \citeasnoun{jagodzinski1954symmetrieeinfluss} it can be shown
\begin{align}
\frac{d\sigma}{d\Omega} =&\frac{\sin^2(N_a\pi h)}{\sin^2(\pi h)}\frac{\sin^2(N_b\pi k)}{\sin^2(\pi k)} Re\bigg\{ \Tr\big( \boldsymbol{\hat{P}}(2\boldsymbol{\hat{S}} + N_c\boldsymbol{I})\boldsymbol{\hat{F}} \big)\bigg\} \label{MNDSCS}
\end{align}
where the terms are defined as follows. First, let $\boldsymbol{Q}$ be the invertible square matrix transforming $\boldsymbol{\Xi}$ into its Jordan Normal form $\boldsymbol{\hat{\Xi}}$ like so
\begin{equation}
\boldsymbol{\Xi} = \boldsymbol{Q}^{-1}\boldsymbol{\hat{\Xi}}\boldsymbol{Q}
\end{equation}
allowing us to define
\begin{equation}
\boldsymbol{\hat{P}} = \boldsymbol{Q}^{-1}\Diag(\boldsymbol{\pi})\boldsymbol{Q} 
\end{equation}
for $\Diag(\boldsymbol{\pi})$ the diagonal matrix with entries the components of the stationary distribution vector $\boldsymbol{\pi}$.
Next, we call $\boldsymbol{F}$ the structure matrix with elements $F_{ij} = F_{i}F^{*}_{j}$ where $F_{i}$ and $F_{i}^{*}$ are the structure factor and conjugate structure factor of an $i$ block respectively. We then let
\begin{equation}
\boldsymbol{\hat{F}} = \boldsymbol{Q}^{-1}\boldsymbol{F}\boldsymbol{Q}
\end{equation}
and note that $\Tr$ denotes the trace. Finally, if $\boldsymbol{\Xi}$ is diagonalisable then $\boldsymbol{\hat{S}}$ is a diagonal matrix with entries
\begin{align}
s_{n} = 
\begin{cases}
\frac{N_c}{2}(N_c - 1) \text{ if $\lambda_n e^{2 \pi i l} = 1$ }  \\
\frac{\lambda_n e^{2 \pi i l} ( \lambda_n^{N_c} e^{2 \pi i l N_c} + N_c(1 - \lambda_n e^{2 \pi i l} ) - 1)}{(1 - \lambda_n e^{2 \pi i l})^2} \text{ otherwise,}
\end{cases}
\label{Sn}
\end{align}
where $\lambda_{n}$ are the eigenvalues shared by the transition matrix $\boldsymbol{\Xi}$ and its diagonal representation $\boldsymbol{\hat{\Xi}}$. On the other hand, if $\boldsymbol{\Xi}$ is not diagonalisable (defective), then $\boldsymbol{\hat{S}}$ is an upper triangular matrix with a more complicated expression. See Appendix \ref{CrossSectionDerivation} for a derivation of equation \eqref{MNDSCS}, where the unlikely and somewhat pathological case of a defective transition matrix $\boldsymbol{\Xi}$ is also covered. By \emph{unlikely}, we mean that almost all transition matrices are diagonalisable, in the measure theoretic sense that almost all real numbers between $0$ and $1$ are not equal to $\frac{1}{2}$. The reason we bother with the defective case at all is to ensure that an optimisation algorithm seeking to find the set of transition probabilites that best fits experimental data would not `break' if the algorithm were to try to evaluate the cross section using a near defective transition matrix; where we informally define a near defective matrix as one for which the process of diagonalisation is numerically unstable. With equation \eqref{MNDSCS} now defined, we use \citeasnoun{PhysRevB.34.3586}'s observation that the cross section of a macroscopic crystal can be well approximated by taking the limit as $N_a$ and $N_b$ tend to infinity, revealing
\begin{align}
\frac{d\sigma}{d\Omega} = &N_{a}N_{b} \delta(h - h_{0})\delta(k - k_{0})Re\bigg\{ \Tr\big( \boldsymbol{\hat{P}}(2\boldsymbol{\hat{S}} + N_c\boldsymbol{I})\boldsymbol{\hat{F}} \big)\bigg\}
\end{align}

We note for experimentalists here that taking a powder average of equation \eqref{MNDSCS} can be problematic near $2\theta = 0$ or $2\theta = \pi$ (measured during backscattering experiments) where broadened Bragg spots cut the \emph{Ewald shell} almost tangentially and are therefore in significant need of a Lorentz correction. On the other hand, the Lorentz correction hardly varies at angles well away form $2\theta = 0$ or $2\theta = \pi$ allowing one to easily bring expression \eqref{MNDSCS} to bear on powder averaged samples for many values of $\theta$.

This expression allows the differential scattering cross section to be computed much more efficiently than a popular method employed by \citeasnoun{PhysRevB.34.3586} to investigate Lithium at 20K, as well as by \citeasnoun{Hansen2008} and by \citeasnoun{kuhs2012extent} to study so-called cubic ice.
The popular method estimates the pair-correlation functions $G_{ij}(m_{3})$, which form part of the Berliner and Werner's expression for the average structure factor product \eqref{SFExpression} and scattering cross section \eqref{BerlinerExpression}. This section compares this method's performance to evaluating expression \eqref{MNDSCS} directly. The popular method begins by sampling block of layers from the crystal. Then a randomly selected layer is fixed atop the seed with probability dependant on preceding $s$ layers, which are initially the set of layers composing the seed. The algorithm continues to grow the crystal by iteratively affixing a new layer atop the last with layer  determined by the previous $s$ layers; this process continues until the crystallite size is of the order of the coherently scattering domain.

The estimate for the pair-correlation $G_{ij}(m_{3})$ is taken by simply counting up the number of times a layer of of type $j$ is found $m_{3}$ layers ahead of one of type $i$, then dividing by the total number of trials.
Since the estimate is reached by sampling a phase space (the set of possible layer sequences) then taking an average to estimate a thermodynamic variable (probability of pair occurrence), this method is a Monte Carlo (MC) simulation. 

\subsection{Monte Carlo vs Markov chain}

The traditional MC algorithm estimates the pair-correlation functions with an uncertainty converging toward $0$ as the number of crystals grown, hence samples taken, tends to infinity. The convergence is asymptotically proportional to the reciprocal square root of the number of samples taken. In big $\mathcal{O}$ notation, the convergence is $\mathcal{O}(N^{-1/2})$ where $N$ is the number of samples \cite{SimTech}. Consequently high accuracy comes at a heavy computational cost, specifically every additional correct decimal point appended to the estimate for the pair-correlation function requires that $100$ times more samples be taken, requiring that the simulation be run for $100$ times longer. Having obtained an estimate for the pair-correlation function, the MC algorithm can compute the scattering cross section at each value of $l$ in $\mathcal{O}(|B|)$ time, where $|B|$ is the number of distinct layer blocks. 
On the other hand, the time taken to evaluate equation \eqref{MNDSCS} for many values of $l$ does not depend on the size of the crystal $N$, instead we compute \eqref{MNDSCS} in $\mathcal{O}(|B|^3)$ for each value of $l$.

Further, since we express \eqref{MNDSCS} exactly, it can be evaluated by a computer as precisely as compounded rounding errors on floating point arithmetic allow. This is deceptively useful because the Levenberg-\citeasnoun{LM}, or similar, algorithm employed by Rietveld's refinement scheme \cite{Rietveld} requires that the sum of square residuals be differentiated with respect to each free parameter and evaluated at every data point for each iteration. The square residual sum is expressed in terms of the cross section, hence Rietveld's method demands that the cross section be differentiated with respect each free parameter including the transition probabilities. Now, it is shown in Appendix \ref{dixB} that analytic derivatives with respect to parameters of the structure factor do exist for \eqref{MNDSCS}, but neither the Monte Carlo algorithm nor \eqref{MNDSCS} easily yield analytic derivatives with respect to the transition probabilities. Consequently, these derivatives must be evaluated numerically using a finite difference. The finite difference approximation evaluates a function at two (or more) points, separated by a small, but finite, length then uses the values to estimate the derivative at a nearby point. Crucially, the approximation is better as the separation between points gets smaller, tending to the exact derivative as the separation tends to $0$. Consequently, computing the cross section to a high precision permits a smaller finite difference, producing a better approximated matrix of derivatives, resulting in a much better behaved descent algorithm.

\subsection{Comparison to the recursion method used by DIFFaX}

\citeasnoun{treacy1991general} have written a user friendly and flexible program in FORTRAN computing the X-ray/neutron scattering pattern of stack faulted crystal structures.  
The program: Diffracted intensities from faulted Xtals (DIFFaX), computes a crystal's differential scattering cross section using an equation that is fundamentally the same as that developed by \citeasnoun{hendricks1942x}, \citeasnoun{warren1959x}, and us in this paper. However, their derivation and heuristic is somewhat different. Roughly speaking, \citeasnoun{treacy1991general} consider that the scattering pattern for a crystal is equal to the pattern of the same crystal shifted upward by one layer superposed with the pattern produced by a single layer placed under the shifted crystal. By relating the scattering pattern to itself recursively, a set of simultaneous equations can set up and solved to yield the scattering pattern of a stack faulted or aperiodic crystal. Solving this set of simultaneous equations has computational complexity $|B|^3$, as does evaluating \eqref{MNDSCS}. However, the time taken to evaluate the exponential function is the most time intensive process if the number of block types $|B|$ is small, which is true for low interaction range and low number of layer types. Though the original DIFFaX does not support a least squares refinement of free parameters, the work in progress DIFFaX+ being developed by \citeasnoun{DIFFaXplus} does support Rietveld refinement.

\section{Close-packed structures}

\subsection{Topology of the close-packed structure}
\label{Topology}
The topologically close-packed aperiodic crystal is ubiquitous in nature and has some interesting mathematical features. First of all, the atomic coordination of a close-packed structure is determined by assuming a crystal layers' constituent atoms are equally sized hard spheres that are physically stacked between other layers. The layers (so-called \emph{modular layers}) are constrained by the structure's geometry to 1 of 3 distinct equilibrium positions labelled $A$, $B$ and $C$ such that adjacent layers cannot have the same relative position. $A$ cannot follow $A$, $B$ cannot follow $B$, nor can $C$ follow $C$. We denote a sequence of layers by printing the positions $A$, $B$ and $C$ and the order these appear in the crystal, exemplified in figure \ref{layerSeq}.

\begin{figure}
\begin{align} 
...ABCABCBACBA... \nonumber
\end{align}
\caption{A subset of a layer sequence.}  \label{layerSeq}
\end{figure}

\begin{figure}
\centering
    \includegraphics[width=0.5\textwidth]{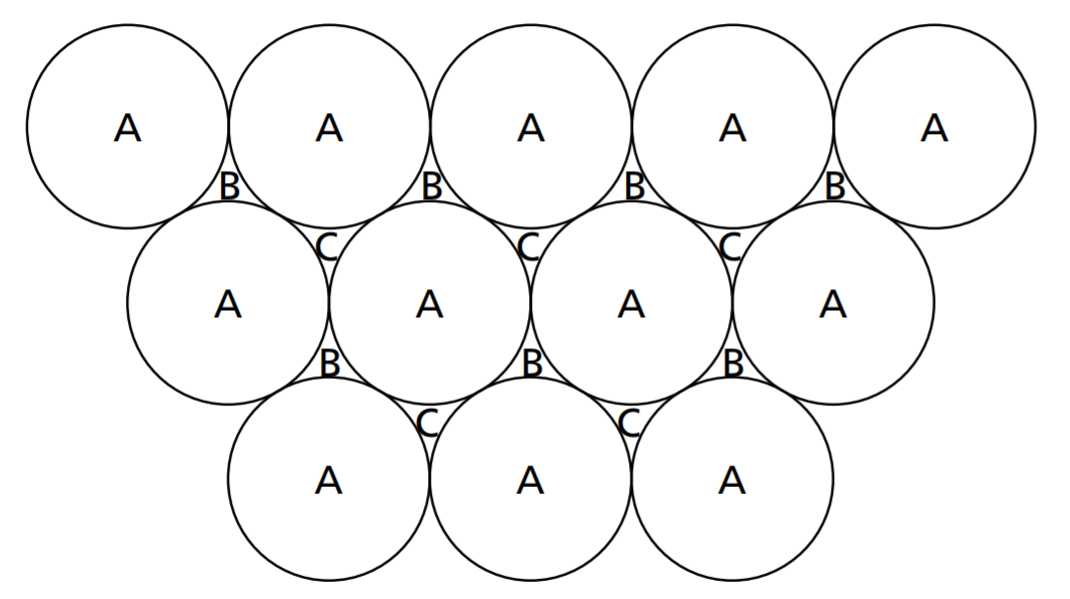}
\caption{Stacking sequences of close-packed layers of atoms. A-first layer (with outlines of atoms
shown); B-second layer; C-third layer. Reproduced from \citeasnoun {FundOfCryst} with permission of the International Union of Crystallography. More images and a short overview of the close-packed structure is provided by \citeasnoun{Closepackedshazzam}.}  \label{diagramOfSpheres}
\end{figure}

It can be seen from figure \ref{diagramOfSpheres} that if a layer is flanked by layers of the same type, the resultant lattice is hexagonal. If it is flanked by layers of different types, the resultant lattice is cubic. We label hexagonally stacked layers $H$ and cubic ones $K$ in accordance with Wyckoff-Jagodzinski notation, then notice all layer sequences have unique representation as a stack sequence. Under this representation, the layer sequence in figure \ref{layerSeq} has a unique stack sequence illustrated and explained in figure \ref{stackSeq}.

We recall that for a 1D aperiodic crystal, a given layers' type depends on finitely many consecutive preceding layers $s$. For a close-packed 1D aperiodic crystal, we observe that the stack type of a layer depends on the stack type of $s-2$ consecutive preceding stacks. For example
a stack sequence with \emph{reichweite} $s$ equal to 4 can be characterised by the transition probabilities $\alpha$, $\beta$, $\gamma$ and $\delta$ that a $K$ stack follows an $HH$, $HK$, $KH$ and $KK$ pair of stacks respectively.

\begin{figure}
\begin{align} 
...KKKKHKKKK... \nonumber
\end{align}
\caption{All layers are flanked by layers of different type, hence they are cubic, except for the layer $C$ at the centre of the subsequence $...BCB...$ which is hexagonal.} \label{stackSeq}
\end{figure}

The Markov process generating the stack sequence is intimately related to that producing the layer sequence. After all, the labels $A$, $B$ and $C$ are deployed arbitrarily to represent a sequence that can be equally well represented in Wyckoff-Jagodzinski $HK$ notation. In order to formally relate these equivalent representations, we first declare that two layer sequences $l_{1}$ and $l_{2}$ belong to the same equivalence class $l_{1} \sim l_{2}$ if and only if there exists a permutation on the set of layer types $\phi$ such that $l_{1} = \phi \circ l_{2}$. We note that all layer sequences belonging to the same class have the same stack sequence representation, consequently, the sequences belonging to the same equivalence class are described by the same set of transition probabilities; hence they are represented by the same Markov chain. Figure \ref{membersOfClass}
displays a stack sequence on its right hand side with members of the corresponding layer sequence equivalence class on its left. We let the crystal's stack block Markov chain have transition matrix $\boldsymbol{X}$ and stationary state vector $\boldsymbol{p}$. 

\begin{figure}
\begin{align}
&...ABCABCABCABC... \nonumber \\
&...ACBACBACBACB... \nonumber \\
&...BACBACBACBAC... \nonumber \\
&...BCABCABCABCA... \nonumber \\
&...CABCABCABCAB... \nonumber \\
&...CBACBACBACBA... \nonumber 
\end{align}
\caption{These six layer sequences belong to the same equivalence class because there is a permutation on the set of layer types that will map any one of the sequences to any other. For example, the permutation $\phi_{1} \colon \{A,B,C\} \to \{A,B,C\}$ defined $\phi_{1}(A) = A, \phi_{1}(B) = C, \phi_{1}(C) = B$ is employed to map the first sequence in the list to the second. This equivalence class comprises all layer sequences with perfectly cubic stack sequence $KKKKKKKKKK$.}  \label{membersOfClass}
\end{figure}

\begin{figure}
    \includegraphics[width=0.5\textwidth]{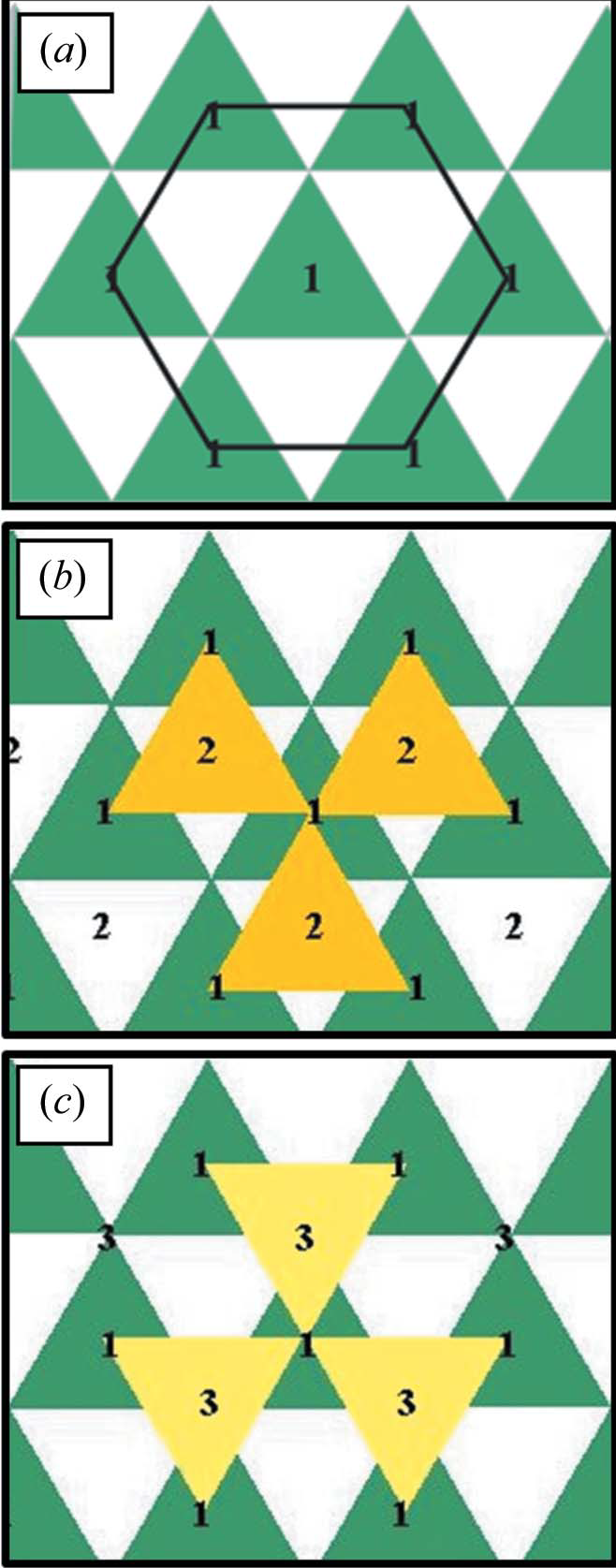}
\caption{Here the labels $1,2,3$ replace $A,B,C$. Notice the arrangement of tetrahedra is the same as the arrangement of spheres dipslayed in figure \ref{diagramOfSpheres}. Reproduced from an educational paper by \citeasnoun{ortiz2013prolific} on the prolific polytypism of silicon carbide, with permission of the International Union of Crystallography.}  \label{tets}
\end{figure}

 Spherically close-packed structures are not the only structures that can be described in $ABC$ or $HK$  notation. For example, 1D aperiodic
 ice crystals are open-packed, but placing hypothetical spheres with appropriate radii at the midpoint of each of an ice crystal's hydrogen bonds results in an ensemble of spheres that is close-packed. Hence, we say that cubic ice is topologically close-packed and describe it using the same $ABC$ or $HK$ notation as real close-packed structures. Similarly, silicon carbide layers are composed of tetrahedra arranged with spherical close-pack topology, as can be seen in figure \ref{tets}. Informally, any aperiodic 1D crystal that can be described in $ABC$ or $HK$ notation has the same spherical close-pack topology.
 
\subsection{The cross section of aperiodic ice}

The simplest close-packed structure is a crystal composed of three layer types $A$, $B$, an $C$ that are identical up to some translation in real space, so their structure factors are equal up to some rotation in reciprocal space, as observed by \citeasnoun{PhysRevB.53.5198}. This structure has been studied extensively by many authors including \citeasnoun{PhysRevB.34.3586}, \citeasnoun{PhysRevB.53.5198}, and \citeasnoun{riechers2014diffraction}, but is insufficient to describe the aperiodic ice studied by \citeasnoun{Hansen2008} and \citeasnoun{hansen2015approximations}. In fact, Hansen \emph{et al.} explain that the content of an ice layer depends on that of its neighbours; so are forced to consider the existence of $6$ distinct layers $AB$, $AC$, $BA$, $BC$, $CA$, and $CB$ each with their own structure factor. These layers are obtained by considering a pair of layers $A$ and $B$ for example, then shifting the borders of the layer $A$ in the aperiodic direction by half a unit cell, obtaining a layer containing half the $A$ layer and half the $B$ layer; then labelling this layer $AB$. This is illustrated in figure \ref{ice_layers_fig}.
\begin{figure}
\begin{align} 
...
\overunderbraces{&\br{2}{AB}}
{ \ B \ \ C \ \ & A \ & \ B \ & \ C \ & \ A \ \ B \ }
{&&\br{2}{BC}} \nonumber 
...
\caption{New layer types from old} 
\end{align}
\label{ice_layers_fig}
\end{figure}
\citeasnoun{Hansen2008} provide complete molecular details including the structure factor of each of the 6 layer types, including the important observation that the layers are not simply identical up to translation. The authors further assume that a layer depends on some finite number $s$ of previous layers, allowing ice to be described as an aperiodic crystal built from a sequence of layer blocks each with length $s$. Under the conditions that $A$ cannot follow $A$, $B$ cannot follow $B$ nor can $C$ follow $C$, ice comprises $|B| = 3 \times 2^s$ distinct block types. The cross section of ice is therefore given by equation \eqref{MNDSCS}, which is in terms of an appropriate transition matrix $\boldsymbol{\Xi}$ and structure matrix $\boldsymbol{F}$. The remainder of this section describes how one might construct these matrices, though before we begin, we note that for low $s$ the resulting matrices are small, and it may be possible to cleverly construct the transition matrix using methods similar to those deployed by \citeasnoun{riechers2015pairwise} to produce analytic expressions for the eigenvalues and cross section.

First of all, we need to define what is meant by \emph{a block of type $i$} when discussing ice. To do this, we index the set of block types with a subset of the integers $\{ 1, \ 2, \ ... \ , 3 \times 2^s \} \ni i$ using the following scheme.  
First, find the quotient $q$ and remainder $r$ upon dividing $i-1$ by $6$. i.e write $i-1 = 6q + r$. The remainder is either $0, 1, 2, 3, 4,$ or $5$; so for each of these numbers respectively, set the first layer in the block as $AB, AC, BA, BC, CA,$ or $CB$. Now express the quotient in binary, producing a sequence of $s-1$ digits that are either $1$ or $0$. Let $X,Y,Z \in \{ A , B , C \}$ and observe that if the $n$th layer of the block
is the layer $XY$, then either it is the last layer of the block, or the next layer is $YZ$ for $Z$ taking one of $2$ values (since $Z \neq Y$). If the $n$th digit of the binary representation of $q$ is $0$ then choose $Z$ depending on $Y$ as follows. If $Y = A$, set $Z = B$, if $Y = B$, set $Z = C$, and if $Y = C$, set $Z = A$. On the other hand if the $n$th digit of the binary representation of $q$ is $1$, then if $Y = A$, set $Z = C$, if $Y = B$, set $Z = A$, and if $Y = C$, set $Z = B$.
This procedure determines the first of a block's layers, then constructs the $n$th layer from the $n-1$th layer, hence recursively constructs the entire block.

This scheme is of course reversible. Indeed the first layer in the block is either $AB, AC, BA, BC, CA,$ or $CB$ so set the remainder $r$ to the respective index $0, 1, 2, 3, 4,$ or $5$ depending on the first layer. Next, define a sequence of $s-1$ binary digits as follows. If the $n+1$th layer in the block is $AB$, $BC$, or $CA$ set the $n$th binary digit to $0$, otherwise, set the digit to $1$. The sequence of digits is a binary representation of a number $q$. With $q$ and $r$ determined, set the index $i = 6q + r + 1$.

So in the context of ice at a given \emph{reichweite}, it is now clear what we mean when referring to a block of type $i$. For example, if $s=4$ then the block of type $21$ is $BA \ AB \ BA \ AC$, which can be seen by noting $21 - 1 = 6 \times 3 + 2 $, so $r = 2$, hence the first of the block's layers is $BA$, and the quotient $3$ has binary representation $011$, fixing the next three layers as $AB$, $BA$, and $AC$.

Next, we need the structure factor of the $i$th block $F_i$ which we can find as follows. 
Let $f_{n}^{i}$ be the structure factor of the $n$th layer in the block of type $i$; which has expression found in the work of \citeasnoun{Hansen2008}. Now a
block's structure factor is just a superposition of the structure factor of each layer shifted to the right position, so 
\begin{align}
F_i = \sum_{n = 0}^{s-1} f_{n}^{i} \exp\bigg(\frac{2 \pi i m_3 l n}{s}\bigg).
\end{align}
Now we can fully define the structure matrix $\boldsymbol{F}$ by recalling its $ij$th elements are just $F_i F_j^*$.

Next on the agenda is to obtain the $ij$th element of the transition matrix. These elements can be discovered by noting that a block $i$ defines a stack sequence $(1)$ in $HK$ notation with length $s-1$. The concatenation of blocks $i$ and $j$ defines a second stack sequence $(2)$ with length $2s - 1$ where the first $s-1$ stacks are those belonging to block $i$. Suppose an arbitrary stack sequence of length $2s - 1$ has its first $s-1$ stacks fixed to those comprising sequence $(1)$, then the probability that the remaining stacks form sequence $(2)$ is the probability that block $i$ is followed by block $j$; which specifies the $ij$th element of the transition matrix.

For example, when $s = 4$ the transition matrix $\boldsymbol{\Xi}$ has element $\xi_{21 \ 35}$ representing the probability that a block of type 35 will follow one of type 21. Using the index scheme, notice this is the probability that the block $CA \ AC \ CA \ AC$ will follow the block $BA \ AB \ BA \ AC$. This is exactly the probability of $ACAC$ following $BABAC$ which equals the probability of $HHHH$ following $HHK$. Recalling that the probability of $K$ following $HH$, $HK$, $KH$, and $KK$ is given by $\alpha$, $\beta$, $\gamma$, and $\delta$ respectively, we have that $\xi_{21 \ 35} = (1-\beta)(1-\gamma)(1-\alpha)^2$.

With the structure matrix $\boldsymbol{F}$ and transition matrix $\boldsymbol{\Xi}$ now defined, computing the cross section of aperiodic ice is just a matter of evaluating equation \eqref{MNDSCS}. It is also possible to compute the entropic density $h$ with only the transition matrix.

\subsection{Aperiodic opal}

There is ample  X-ray diffraction data \cite{Graetsch1994},\cite{GDGuthrie} to suggest that the opal CT form of silica is composed of disordered (cubic) cristobalite and (hexagonal)
tridymite layers, in topological identity to disordered ice. Like ice, these opaline forms of silica are topologically close-packed
stackings of $ABC$ (cristobalite) and $AB$ layers (tridymite), in which the packing centers are located at
the linear midpoint of the Si-O-Si covalent bond chains, i.e. close to the (likely disordered) oxygen
position. This view is supported by HRTEM studies in which the stacking disorder could
be directly visualised \cite{JMElzea}. It was also recognised that the crystallite sizes of the
stacking disordered opals is nanoscopic \cite{GDGuthrie} leading to diffraction broadening. It
was further suggested that additionally disordered, more amorphous regions could be an intrinsic part
of opal CT. The various micro-structural aspects of opal CT were recently discussed by \citeasnoun{WILSON201468},
who highlighted some discrepancies between diffraction and spectroscopic findings. As suspected by
\citeasnoun{GDGuthrie}, a straightforward assignment of cubic and hexagonal peak intensities in the
complex first diffraction (triplet) peak to the relative proportions of tridymite and cristobalite layers
seems unjustified; a statement supported by the work of \citeasnoun{JMElzea}. 

\citeasnoun{Arasuna2013} have suggested that a largely amorphous water-containing opal (so-called
opal A) that undergoes heating and annealing will transform (under loss of water) continuously into a progressively more crystalline opal CT form, finally becoming a material dominated by cristobalite stackings; though the authors adopt a simplified view of stacking disorder which is unlikely to be
quantitatively correct. A more involved treatment of stacking disorder and micro-crystallinity as
presented here and in the past for ice \cite{Hansen2008} \cite{kuhs2012extent} has not yet been applied to opal but appears highly desirable as it may well resolve some of the open issues on the nature
of disorder in opal CT as discussed by \citeasnoun{WILSON201468}. In any case, the annealing of amorphous silica
via stacking disordered opal CT into a largely crystalline form close to the melting point of silica
shows a close resemblance to the annealing of amorphous ice with the main difference that opals
drive towards a cubic form close to melting while ice prefers a hexagonal arrangement.

It is also noteworthy that \citeasnoun{GDGuthrie} were interested in the X-ray diffraction pattern of opal containing water molecules. Our formalism could capture this by letting an opal unit cell containing an H$_2$O molecule at a particular position and orientation within the cell be a new layer type with appropriate structure factor.

\subsection{Reversible crystals}
\label{ReversibleCrystals}

We are interested in whether a 1D aperiodic crystal is reversible, which we define informally as whether it looks the same (in some statistical sense) upside down. 
Before proceeding, we should declare that we have appropriated the word \emph{reversible} from Markov theory and do not mean it in the thermodynamic sense of a process maintaining a constant entropy. We also restrict ourselves to the special case of a topologically close-packed 1D aperiodic crystal in order to conveniently relate this theory to the experimental findings of \citeasnoun{Hansen2008} \& (2008$b$); but note that the idea of a reversible crystal is quite general. Further, we restrict ourselves to crystals with layer types that are individually inversion symmetric. Next, we observe that since a set of layer chain equivalence classes is related to a unique stack chain, we say that a layer chain equivalence class contains reversible layer chains if and only if its related stack chain is reversible.

With the preamble out of the way, we define a reversible crystal as one for which the probability of sampling a type $i$ block of stacks and discovering the next block is type $j$ matches the odds of sampling from the crystal a block with stack sequence the reverse of $j$, and noting its successor is the block type with stack sequence reversing that of block type $i$. In other words, sampling a two block long sequence of stacks running from the beginning of an $i$ block's sequence to the end of a $j$ block's is as likely as the sequence running from the end of the $j$ block's to the beginning to the $i$ block's.
Formally, an aperiodic crystal is reversible if and only if its underlying Markov process satisfies the reversibility condition
\begin{equation}
p_{i}X_{ij} = p_{\sigma(j)}X_{\sigma(j)\sigma(i)} \ \forall \ ij \label{ReversibilityCondition}
\end{equation}
where $\sigma \colon B \to B$ is the involution mapping a block type index to the index of the block type with the reverse stack sequence. Note that we are not using Einstein's sum notation; reversibility entails that equation \eqref{ReversibilityCondition} holds pointwise over all $i$ and $j$.

We are interested in whether the $s = 4$ close-packed 1D aperiodic crystal in particular is reversible, so we note first of all that the reversibility condition holds trivially for $ij$ satisfying $\sigma(i) = j$. Next, using that this crystal's involution map is defined $\sigma(1) = 1, \sigma(2) = 3, \sigma(3) = 2, \sigma(4) = 4$ we see that 
\begin{align}
p_{1}X_{12} &= (1 - \gamma)(1 - \alpha)p_{2} &= p_{3}X_{31} = p_{\sigma(2)}X_{\sigma(2)\sigma(1)} \\
p_{1}X_{13} &= (1 - \gamma)(1 - \beta)p_{2} &= p_{2}X_{21} = p_{\sigma(3)}X_{\sigma(3)\sigma(1)}\nonumber \\
p_{1}X_{14} &= (1 - \gamma)\beta p_{2} &= p_{4}X_{41} = p_{\sigma(4)}X_{\sigma(4)\sigma(1)}\nonumber \\
p_{2}X_{22} &= (1-\beta)\gamma p_{2} &= p_{3}X_{33} = p_{\sigma(2)}X_{\sigma(2)\sigma(2)}\nonumber \\
p_{2}X_{24} &= \beta\delta p_{2} &= p_{4}X_{43} = p_{\sigma(4)}X_{\sigma(4)\sigma(2)}\nonumber \\
p_{3}X_{34} &= \gamma\beta p_{2} &= p_{4}X_{42} = p_{\sigma(4)}X_{\sigma(4)\sigma(3)}\nonumber 
\end{align}
revealing the remarkable fact that all $s = 4$ close-packed 1D aperiodic crystals are reversible. Notice that if we let $\alpha = \gamma$ and $\beta = \delta$ the probability that a stack is type $K$ depends only on a single preceding stack, which is to say $s = 3$, and the crystal is still reversible. Now if we let $\alpha = \beta = \gamma = \delta$ we are left with a sequence of independent and identically distributed random variables where $s = 2$ and that the crystal is of course still reversible. Thus we conclude that the crystal is reversible for all $1 < s \leq 4$.

This result does not hold for crystals where $s > 4$. Specifically, for crystals with \emph{reichweite} greater than 4 some sets of transition probabilities ($\alpha, \beta, \gamma$ etc) satisfy condition \eqref{ReversibilityCondition} while others do not. We can see that there exist reversible crystals with $s>4$ by considering a chain of blocks where a block composed entirely of cubic stacks will certainly follow one composed entirely of hexagonal stacks, and vice versa. This reversible crystal happens to possess the property of being a polytype, and can be shown to satisfy the reversibility condition. Next, we claim that when $s > 4$ some chains are irreversible, which we prove with the following example. Let $s = 5$ and the probability of $K$ following
$HHH \text{ be } 1, \ $
$HHK \text{ be } 1, \ $
$HKK \text{ be } 0, \ $
$KKH \text{ be } 1, \ $
$KHK \text{ be } 0 $
and the probability be $\frac{1}{2}$ for all other blocks. Note that the blocks $HHH, \ KKH, \ KHH, \ HKK, \ HKH, \ HHK, \ KHK $ form a closed communicating class where the probability of one following the other is either $0$ or $1$, so represents a polytype with stack sequence representation shown in figure \ref{Irreversible}. The polytype does not satisfy the reversibility condition and is therefore irreversible. 
\begin{figure}
\begin{align} 
...\underbrace{HHHKKHK}_{\text{polytypic lattice}}\underbrace{HHHKKHK}_{\text{polytypic lattice}}... \nonumber
\end{align}
\caption{This crystal is irreversible because the block $HHH$ is certainly followed the block $KKH$, but in reverse $HHH$ is certainly followed by $KHK$.}  \label{Irreversible}
\end{figure}
This result can be extended to any $s > 4$ by simply considering a crystal composed of lattices like those in figure \ref{Irreversible} but with more than $3H$ stacks in a row before being followed by the sequence $KKHK$. In summary, we
have established that for a the topologically close-packed crystal with symmetric layer types, if $s \leq 4$ the crystal is certainly
reversible. If $s > 4$ the transition probabilities may or may not satisfy the reversibility condition, hence
the crystal may or may not be reversible. 

Having established that existence of irreversible close-packed crystals with $s > 4$ we are interested in how many reversible crystals can
possibly exist in comparison to irreversible crystals. For $s \leq 4$ we have established the number of irreversible crystals is zero, so we direct our
attention to the case where $s > 4$. A reversible $s > 4$ aperiodic crystal has transition probabilities
constrained by the reversibility condition. This constraint is holonomic, so by the implicit function
theorem, the space of transition probabilities for which a crystal is reversible has lower dimension
than the space of all transition probabilities, so has $0$-measure.
Therefore, reversible crystals with $s > 4$ almost surely do not occur at random, so any that appear in nature almost surely owe their reversibility to some long range interaction somehow forcing them to be symmetric, as it has been observed for polytypic stacking disorder in SiC \cite{varn2001crystal}. 

Finally, we note that the implications of whether a Markov model underlying a dynamical system is reversible has many more subtle and interesting points elaborated at length by \citeasnoun{Ellison2009}.

\subsection{Markov processes in relation to growth physics of stacking disordered ice}

The Markov theoretic considerations in the previous chapter may shed some light on the process by which some class of crystals form. In particular, if an aperiodic crystal is irreversible, then its
formation process must have an intrinsic directionality. Contrapositively, any formation process that
does not have any particular or special directionality should produce a reversible aperiodic crystal.

This observation can be applied to ice, sometimes called I$_{\text{ch}}$ \cite{hansen2015approximations} or ice I$_{\text{sd}}$ \cite{Malkin1041}, a material which can be well described without requiring a \emph{reichweite} $s > 4$ \cite{Hansen2008}, for which there is no recent evidence for the existence of polytypes \cite{Hansen2008}; an earlier specific search for polytypes was similarly fruitless \cite{Kuhs1987}.
Having no experimental evidence for growth processes with $s > 4$ likely means
that the growth of ice stacks is an intrinsically symmetric process; this is in contrast to other materials
with longer-ranged or even infinite \emph{reichweite} discussed by \citeasnoun{varn2016did}. Can we learn something else for the growth physics from a Markov theoretic description of 1D periodic crystals? First of all, the stacking disorder in ice I$_{\text{ch}}$ is very strongly
influenced by the parent phase, both in its \emph{reichweite} and in the frequency of the
distinguishable stackings observed; indeed, the stacking disorder provides very clear and
reproducible fingerprints to trace back the parent phase after its transformation into ice I$_{\text{ch}}$ \cite{kuhs2012extent}. This information is wiped out only upon prolonged annealing \cite{0953-8984-20-28-285105}, \cite{kuhs2012extent}; this is
understood to be a consequence of annihilation of various partial dislocations \cite{doi:10.1080/14786435.2015.1091109}, a process
which also depends on the lateral extent of the stacks. This process proceeds in a discontinuous
manner and eventually yields good hexagonal ice on approaching 240K within a laboratory
timescales of seconds to minutes \cite{B412866D} in full agreement with \citeasnoun{doi:10.1080/14786435.2015.1091109}.
Interestingly, a satisfactory description of ice I$_{\text{ch}}$ obtained from high pressure ices (recovered to ambient pressure at low
temperature) or obtained from water vapour requires $s=4$, making use of 4 parameters $\alpha, \beta, \gamma$ and $\delta$ \cite{kuhs2012extent}, but the formation from super-cooled water
is adequately described by $s=2$, making use of single stacking fault parameter \cite{Malkin1041} \cite{doi:10.1021/acs.jpclett.7b01142}. An explanation for this difference is certainly worth pursuing.

\citeasnoun{kuhs2012extent} introduced the term \emph{cubicity} to describe the proportion of cubic
sequences in a crystal, which has been found to be almost 80\% when freezing very small (15 nm) droplets at
225K \cite{doi:10.1021/acs.jpclett.7b01142}, while larger (900 nm) drops freeze to a 50\%:50\% mixture of cubic and hexagonal
sequences at 232K \cite{Malkin1041}. It turns out that highest cubicities are obtained when no time is allowed for any
annealing of stacking faults, like in the very fast (timescale of $\mu$s) freezing achieved by \citeasnoun{doi:10.1021/acs.jpclett.7b01142}.
This poses a question on the nature of the initially formed nucleus: is it cubic, hexagonal or
stacking-disordered? While the bulk crystal in its stable form is certainly hexagonal, the reasons for
this preference are somewhat less clear: On the basis of quantum mechanical calculations \citeasnoun{PhysRevX.5.021033} have suggested that the anharmonic vibrational energies favour the hexagonal form as a
consequence of differences in the fourth nearest-neighbour protons, related to the occurrence of the topologically different boat and chair-forms of the 6-membered water rings in cubic and hexagonal ice. Still, as
nucleation (and growth) for super-cooled water is kinetically controlled, freezing may well start also
with a cubic or a stacking disordered nucleus \cite{Lupi2017}. The subsequent growth appears experimentally to
follow a fast $s=2$ Markov chain prescription, before any annealing has time to set in
for temperatures below approximately $240$K. It is noteworthy that branching (twinning) has been observed for a
significant percentage of snow crystals formed at temperatures higher than about 238K \cite{1980416}, which likely have formed from a cubic (or stacking-faulty) nucleus growing along directions
separated by the octahedral angle of $70.53^{\circ}$ (i.e. the angle between two or more cubic [111]
directions). Maintaining a larger macroscopic snow-crystal in a stacking disordered state is
energetically expensive due to the development of large-angle grain boundaries between several stacking
directions \cite{Kobayashi1987}; so individual branches develop by further growth
from the gas phase after the initial freezing. Moreover, at high enough temperatures the stacking
disorder will quickly anneal as discussed above. Thus, the only traces left of the earlier stacking
disorder are these multiply twinned, branched hexagonal crystals.

But why is vapour-grown ice I$_\text{ch}$ so complex that a satisfactory description demands that $s=4$? The
observed preferential stacking sequences \cite{kuhs2012extent} indicate a persistence of
hexagonal or cubic consecutive stackings rather than a frequent switching between them and an
overall preponderance of hexagonal stackings. Such a persistence can easily be explained by the
growth of stackings around screw dislocations \cite{Thrmer11757}, a growth mechanism which
avoids a costly layer-by-layer nucleation along the stack.
Recovered high-pressure phases of ice (like ice IX or ice V) when transformed into ice I$_{\text{ch}}$ do not show
strong indications for persistence nor for alternating stackings. Rather, the stackings developed could
well reflect orientational relationships with their parent phase. Such a topological inheritance has
been demonstrated in the ice I${_\text{h}}$ $\to$ ice II transition \cite{doi:10.1080/01418619708209983} and is manifest in the observed textural
relationships. It is well conceivable that structural inheritance could also express itself in certain
stacking topologies to minimize the bond-breaking as well as striving for the shortest pathways for
(multistage) diffusionless reconstructive phase transitions \cite{Christy:al0563}.

It is also worth mentioning that the topology of the ice IX phase is acentric, with water molecules arranged along a 4-fold screw-axis. In particular, there are two enantiomorphic forms of ice IX (and the same is true for ice III) with left and right handed forms occurring in nature with equal probability. Consequently, a naturally occurring sample of ice IX is expected have equal proportions of right and left handed crystallites. Now, we expect a sequence of right handed layers to be the reverse of a sequence of left handed layers, and since we expect a sample to contain both forms in equal proportions, an irreversible ice IX crystal would not be simply distinguished from a reversible counterpart by examining their X-ray or neutron powder diffraction patterns.

Further work is undoubtedly needed to elucidate
the myriad of transitions between the many forms of ice in search for an explanation for the observed stacking probabilities.

\section{The pair distribution function of a 1D aperiodic crystal}

The scattering-length density function $\beta(\vec{r})$ describes the distribution of scatterers of the ensemble when centred at the origin \cite{sivia2011elementary}. The autocorrelation function of $\beta(\vec{r})$, which is given by its convolution with its complex conjugate, produces a pair-correlation function that we will denote $g(\vec{r})$.  There are several expressions for pair-correlation functions that differ in their normalisations, and they can be written either in vectorial or in orientationally-averaged form. See \citeasnoun{fischer2005neutron} and \citeasnoun{keen2001comparison}.  

The pair-correlation function $g(\vec{r})$ of an aperiodic ensemble of scatterers (atoms) gives the probability
density of finding an atom a vector distance $\vec{r}$ from an ensemble-averaged atom at the origin. It can be obtained by Fourier transforming the total scattered intensity $I(\vec{Q})$ and is usually separated into a (trivial) self-correlation part and a structure-dependent so-called distinct part
\begin{equation}
g(\vec{r})_{self} + g(\vec{r})_{distinct} \propto \int_{Q-\text{space}} I(\vec{Q}) e^{-2\pi i \vec{Q}\cdot\vec{r}} d\vec{Q}.
\end{equation}
$I(\vec{Q})$ is obtained experimentally as the total differential scattering cross section into solid angle $d\Omega$ as
e.g. measured in a diffraction experiment; this measured total intensity is composed of the trivial self-scattering part and the structurally more interesting distinct part: 
\begin{equation}
I(\vec{Q}) = \frac{d\sigma(\vec{Q})}{d \Omega} = I(\vec{Q})_{self} + I(\vec{Q  })_{distinct}.
\end{equation}
For powders with a random orientation of particles, the diffraction data obtained are usually 1D
averages of $I(\vec{Q})$ like those obtained for amorphous materials or liquids; consequently only
the isotropic function $g(r)$ can be accessed experimentally. Yet, for known atomic arrangements
of the powder crystallites the isotropic average of their 3D pair-correlation functions can be
obtained by integration over the 3D shell at constant $r$. The isotropic $g(r)$, which by choice contains only the distinct part, can then be renormalised into a Radial Distribution Function or RDF$(r)$ from which coordination numbers can be obtained by integration. It is also possible to normalise $g(r)$ into the density function $D(r)$ whose slope at small $r$ is proportional to the sample's atomic number density. It is this function $D(r)$, when generalised to polyatomic systems, that is frequently called the Pair Distribution Function, or PDF$(r)$.  Note that the PDF$(r)$ converges to zero at large $r$, since it represents fluctuations around the average atomic density.

PDF-analysis has developed into an important tool for analysing the often defective
atomic arrangements of nanomaterials \cite{neder2014pdf}, \cite{egami2012underneath}. Nanocrystalline materials often exhibit
stacking-faults as a consequence of their manufacturing procedures (ball milling, mechanical
alloying) or crystal growth \cite{zehetbauer2009bulk}. Such materials show very broad
reflections as a consequence of crystal size broadening and stacking-faults as well as microstrains, and
thus are not routinely accessible by Rietveld analyses if these contributions are not
disentangled \cite{gayle1995stacking}. A PDF-analysis of stacking-faults in nanomaterials seems a
viable alternative and was performed e.g. for CdSe by \citeasnoun{yang2013confirmation} as well as by \citeasnoun{masadeh2007quantitative} and \citeasnoun{gawai2016study} for ZnS nanocubes and nanowires. The stacking-fault model
used in these works is rather simple and limited to mixture models of the pure cubic and
hexagonal constituents via a single stacking-fault probability parameter. That said, a more sophisticated treatment of disordered ZnS was presented by \citeasnoun{PhysRevB.66.174110} who offer a broad statistical description of the close-packed topology, which includes ZnS. The authors find the minimum
effective memory length for stacking sequences in close-packed structures and discuss how to infer the $\varepsilon$-machine from scattering data.

The pair distribution function $g(r)$ is sensitive to the
next-nearest neighbour arrangements, consequently also to the \emph{reichweite} of the layer
interactions, it is in principle possible to extract detailed information on the nature of stacking
faults from an experimental $g(r)$ using a Markov chain approach. 
Obtaining $g(r)$ via a Fourier transformation of the (incompletely and imperfectly)
measured $I_{meas}(Q)$ results in noise and artefacts, so it might well be worth calculating $g(r)$ from the direct-space structure model scattered intensity $I_{model}(Q)$, accounting for instrument resolution etc, then comparing with $I_{meas}(Q)$, as has been done for nanocrystalline, stacking-faulty ice by \citeasnoun{Hansen2008}, (2008$b$), and \citeasnoun{kuhs2012extent}.
Indeed, a $Q$-space based approach like Rietveld refinement is the only way to proceed in cases where high $Q$ data are not
available for making a meaningful Fourier transformation to obtain $g(r)$ from $I(Q)$. 

We should stress that neither $Q$-space nor PDF-analysis is generally better than the other,
but that they are chosen carefully depending on different experimental situations:
 For example, a low density of defects in an otherwise crystalline system is better analysed using $Q$-space refinement since $S(Q)$ displays long-range correlations of defects as diffuse scattering near the base of Bragg peaks. On the other hand, for a high density of defects, especially when one begins to see broad and/or asymmetric intrinsic profiles of Bragg peaks, which can
also happen for quasi-2D or quasi-1D systems, then PDF-analysis is in principle better than
Rietveld refinement. In these cases, the methods of efficiently calculating PDFs of aperiodic crystals developed by \citeasnoun{varn2013machine} and
\citeasnoun{riechers2015pairwise} may come in handy. Moreover, the resolution of the neutron diffractometer plays a big role in deciding
between PDF and $Q$-space analysis, at least for reactor-based diffractometers where there is a tradeoff between high $Q$ (needed for good PDF-analysis and
resolution in R-space), and good $Q$-space resolution (needed for seeing
diffuse scattering at the base of Bragg peaks). 
It is only with some spallation-source diffractometers that one can
achieve very good $Q$-space resolution in addition to a high $Q_{\text{max}}$
(at the expense of counting rate at high $Q$), and in that case one
could conceivably attempt both PDF and $Q$-space analysis.

\section{Conclusion and outlook}

We have mildly generalised the cross section derived by \citeasnoun{riechers2015pairwise} to reach an expression for the scattering cross section of crystals including ice and opal. These crystals have close-packed topology, so we studied the close-packed topology in more detail, finding that those topologically close-packed crystals with \emph{reichweite} 4 or less are necessarily reversible and those with \emph{reichweite} greater than 4 are almost surely irreversible.

Our expression for the cross section provides the experimental crystallographer with a description that fully accounts for the stacking disorder of ice and opal (and possibly more) when applying a Rietveld-like analysis. This could be used to estimate transition probabilities, and understand the distribution of stacking faults much more accurately than trying to estimate them via MC simulation. Moreover, one could seek to determine the \emph{reichweite} of opal in the same way that \citeasnoun{Hansen2008} and \citeasnoun{kuhs2012extent} measured the \emph{reichweite} of ice, which was fundamentally similar to the method outlined by \citeasnoun{PhysRevB.66.174110} to determine the $\varepsilon$-machine of a close-packed crystal. Roughly speaking both approaches involve attempting to fit a model with $s = 1$ to scattering data, and if the model fit is not \emph{good enough} by some metric, the methods increment $s$ until the fit becomes good enough; though the $\varepsilon$-machine reconstruction of \citeasnoun{PhysRevB.66.174110} is somewhat more sophisticated. Further to this, one could examine how the transition probabilities and entropic density of opal evolve under change in temperature, or any other variable. Obtaining the \emph{reichweite} of opal could provide information about its reversibility, providing clues about its formation process.

A fruitful direction of future theoretical work may be to extend the theory so far explored to crystals composed of infinitely many kinds of layer, which could be applied to a crystal composed of layers that are identical to their immediate predecessor up to some rotation, translation, change in curvature, or shift orthogonal to the basel plane, which takes one of infinitely many values. Such crystals possess so-called turbostratic disorder, and include a range of materials including smectites \cite{Ufer:2008:0009-8604:272}, \cite{Turbostratic2009}, carbon blacks \cite{ShiThesis} \cite{ZHOU201417} and possibly $n$-layer graphene; a novel material that has captured the attention of the nanoscience community \cite{Razado-Colambo2016} \cite{Huang2017}. Such an extension of existing theory may be achievable by replacing the transition matrix (an operator on a finite dimensional vector space) with a transition operator on an infinite dimensional Banach space. This functional analytic treatment could be extended to hidden Markov models with infinitely many alphabetical symbols as well as an infinite state space.
\\

\ack{We thank the Institut Laue Langevin for funding A. G. Hart's internship. Further, we thank Joellen Preece for advice on Markov theory and probability, as well as Henry Fischer for guidance on PDF analysis. Further thanks are owed to the anonymous reviewers for their knowledgeable and detailed suggestions which helped to considerably improve the manuscript.  We extend our gratitude to Chris Cook, Michael Green, Matthew Hill, Daniel Hoare, and Lucy Roche for offering corrections and criticism.}

\begin{appendix}

\section{Cross section of a 1D aperiodic crystal described by a HMM}
\label{HMM_CS}

HMMs are an ordered quintuple $\Gamma = (A,\mathbb{S},\mu_0,\mathcal{T},V)$ where
$\mathbb{S}$ is the state space of some hidden Markov process giving rise to a sequence of states $\mathcal{S}_n$ satisfying the Markov property. $\mathcal{T}$ is the transition matrix between states of the hidden Markov process, with elements $\mathcal{T}_{rs}$ representing the probability that a hidden state $r \in \mathbb{S}$ will transition to another hidden state $s \in \mathbb{S}$. $\mu_0$ is some initial probability distribution over the state space $\mathbb{S}$. $\mathcal{A}$ is the alphabet of symbols, which are not hidden, and represent the set of distinct layer types. At every state $s \in \mathbb{S}$ some symbol from the alphabet $\mathcal{A}$ will be emitted with probability following a distribution dependant only on $s$. Specifically, the probability that the state $s \in \mathbb{S}$ will emit a symbol $x \in \mathcal{A}$ is the element $V_{sx}$ of the matrix $V$. The sequence of hidden states $\mathcal{S}_n$ gives rise to a sequence of symbols $X_n$; which is of course the sequence of layer types that compose a crystal.

Our definition of a HMM is presented differently to that of \citeasnoun{riechers2015pairwise}, but is equivalent. In fact, we can define a set of matrices $\boldsymbol{T}$ with elements $\mathcal{T}^{[x]}$ for each $x \in \mathcal{A}$ with components $\mathcal{T}^{[x]}_{rs} = \mathcal{T}_{rs}V_{sx}$ representing the probability of both a transition to state $s \in \mathbb{S}$ from state $r \in \mathbb{S}$ and an emission of symbol $x \in \mathcal{A}$ from state $s$. Then we recover the quadruple $(A,\mathbb{S},\mu_0,\boldsymbol{T})$ used by \citeasnoun{riechers2015pairwise} to define the HMM.

Now, to derive the cross section we first consider the average structure factor product $Y_m$ expressed by \citeasnoun{PhysRevB.34.3586} in terms of structure factors and pair correlation functions of the layer types comprising a crystal. To this end we let $F_{x}$ represent the structure factor of the layer type $x \in \mathcal{A}$, and $F$ be a matrix with entries $F_{xy} = F_{x}F^*_{y}$. Further we let $\pi$ represent the stationary distribution of the hidden Markov process, which exists if the hidden Markov process is positive recurrent and irreducible. Starting from the average structure factor product $Y_m$ provided by \citeasnoun{PhysRevB.34.3586}, making use of Bayes' theorem and the Markov property of the sequence $\mathcal{S}_m$, we have for $m > 0$

\begin{align}
    Y_{m} = 
    \sum_{x \in \mathcal{A}} \sum_{y \in \mathcal{A}}P(X_0 = x , X_{m} = y) F_{xy} \\ =
    \sum_{x \in \mathcal{A}} \sum_{y \in \mathcal{A}}P(X_0 = x)P(X_m  = y \mid X_0 = x) F_{xy} \\ =
    \sum_{x \in \mathcal{A}} \sum_{y \in \mathcal{A}}\sum_{r \in \mathbb{S}} \sum_{s \in \mathbb{S}} P(\mathcal{S}_0 = r)P(X_0 = x \mid \mathcal{S}_0 = r) \\ \times P(\mathcal{S}_m = s \mid \mathcal{S}_0 = r)P(X_m = y \mid \mathcal{S}_m = s)F_{xy} \nonumber \\ =
     \sum_{x \in \mathcal{A}} \sum_{y \in \mathcal{A}}\sum_{r \in \mathbb{S}} \sum_{s \in \mathbb{S}} \pi_s v_{rx} \mathcal{T}_{sr}^m v_{sy} F_{xy} \\ =
     \sum_{x \in \mathcal{A}} \sum_{y \in \mathcal{A}}\sum_{r \in \mathbb{S}} \sum_{s \in \mathbb{S}} \pi_s  \mathcal{T}_{sr}^m v_{rx} F_{xy} v_{sy} \\ =
     \Tr\big( \Diag(\pi) \mathcal{T}^m V F V^{T} \big).
\end{align}
Now,
\begin{align}
    Y_{-m} = \sum_{x \in \mathcal{A}} \sum_{y \in \mathcal{A}}P(X_0 = x , X_{m} = y) F_{yx} \\ =
     \sum_{x \in \mathcal{A}} \sum_{y \in \mathcal{A}}P(X_0 = x , X_{m} = y) F_{xy}^*
\end{align}
because $F$ is Hermitian, so
\begin{align}
    Y_{-m} = \Tr\big( \Diag(\pi) \mathcal{T}^m V F^* V^{T} \big),
\end{align}
and last of all
\begin{align}
Y_0 = \sum_{x \in \mathcal{A}} \sum_{y \in \mathcal{A}}P(X_0 = x , X_0 = y) F_{xy} \\ =
\sum_{x \in \mathcal{A}} \sum_{y \in \mathcal{A}}\pi_s \delta_{sr} v_{rx} F_{xy} v_{sy} \\ =
\Tr\big( \Diag(\pi) V F V^T \big).
\end{align}
Next,
\begin{align}
    \sum_{m = 1}^{N} (N - |m|) Y_m e^{2 \pi i m l} \\ = \Tr\bigg( \Diag(\pi) \sum_{m = 1}^{N} (N-m) \big(\mathcal{T} e^{2 \pi i l}\big)^m V F V^{T} \bigg) \nonumber \\ =
    \Tr\bigg( \Diag(\pi) S V F V^{T} \bigg)
\end{align}
where
\begin{align}
    S = \sum_{m = 1}^{N} (N-m) \big(\mathcal{T} e^{2 \pi i l}\big)^m \\ =
    \mathcal{T} e^{2 \pi i l} ( \mathcal{T}^{N} e^{2 \pi i l N} + N(I - \mathcal{T} e^{2 \pi i l} ) - I)
    (I - \mathcal{T} e^{2 \pi i l})^{-2} \nonumber
\end{align}
if $(I - \mathcal{T} e^{2 \pi i l})^{-1}$ exists. Similarly 
\begin{align}
    \sum_{m = -N}^{-1} (N - |m|) Y_m e^{2 \pi i m l} \\ = \sum_{m = 1}^{N} (N - m) Y_{-m} e^{- 2 \pi i m l} \nonumber \\ =
    \Tr\bigg( \Diag(\pi) S^* V F^* V^{T} \bigg)
\end{align}
so
\begin{align}
    \sum_{m = -N}^{N} (N - |m|) Y_m e^{2 \pi i m l} \\
    = \Tr\bigg( \Diag(\pi) S V F V^{T} \bigg) + \Tr\bigg( \Diag(\pi) S^* V F^* V^{T} \bigg) \\ + \Tr\bigg( \Diag(\pi) V F V^T \bigg) \nonumber \\ = 
    2 Re \bigg\{ \Tr\big( \Diag(\pi) S V F V^{T} \big) \bigg\} + \Tr\big( \Diag(\pi) V F V^T \big).
\end{align}
The general expression for the cross section given by \citeasnoun{PhysRevB.34.3586}
\begin{align}
    \frac{d\sigma}{d\Omega} =\frac{\sin^2(N_a\pi h)}{\sin^2(\pi h)}\frac{\sin^2(N_b\pi k)}{\sin^2(\pi k)} \sum_{m_3 = -N_c}^{N_c} (N_c - |m_3|) Y_{m_3} e^{2 \pi i m_3 l}
\end{align}
completes the derivation
\begin{align}
    \frac{d\sigma}{d\Omega} =\frac{\sin^2(N_a\pi h)}{\sin^2(\pi h)}\frac{\sin^2(N_b\pi k)}{\sin^2(\pi k)} \\ \times
    \bigg( 2 Re \bigg\{ \Tr\big( \Diag(\pi) S V F V^{T} \big) \bigg\} + N_c\Tr\big( \Diag(\pi) V F V^T \big) \bigg) \nonumber \\
    =\frac{\sin^2(N_a\pi h)}{\sin^2(\pi h)}\frac{\sin^2(N_b\pi k)}{\sin^2(\pi k)}
    Re \bigg\{ \Tr\big( \Diag(\pi) (2S+N_cI) V F V^{T} \big) \bigg\}.
\end{align}

\section{Expressing the differential scattering cross section of 1D aperiodic crystals}
\label{CrossSectionDerivation}

We begin with the expression for the differential scattering cross developed by \citeasnoun{PhysRevB.34.3586}
\begin{align}
&\frac{d\sigma}{d\Omega} = \frac{\sin^2(N_a\pi h)}{\sin^2(\pi h)}\frac{\sin^2(N_b\pi k)}{\sin^2(\pi k)} \times \label{BerlinerExpressionAppendix} \\
&\sum^{N_{c}}_{m_{3} = -N_{c}}(N_{c} - \abs{m_{3}})Y_{m_3}e^{2 \pi i m_{3} l} \nonumber
\end{align}
and proceed by splitting the sum
\begin{align}
&\sum^{N_{c}}_{m_{3} = -N_{c}}(N_{c} - \abs{m_{3}})Y_{m_3}e^{2 \pi i m_{3} l} \label{blehh} \\
=Y_{0}N_{c} + &\sum^{N_{c}}_{m_{3} = 1} (N_{c} - m_{3})Y_{m_3}e^{2 \pi i m_{3} l} \nonumber \\
+ &\sum^{N_{c}}_{m_{3} = 1} (N_{c} - m_{3})Y_{-m_3}e^{-2 \pi i m_{3} l}. \nonumber 
\end{align}
Now, according to \citeasnoun{PhysRevB.34.3586}, the average structure factor product has expression
\begin{equation}
    Y_{m_{3}}(\vec{Q}) = 
    \sum_{i \in B}\sum_{j \in B} F_{i}(\vec{Q})F_{j}^{*}(\vec{Q   })G_{ij}(m_{3})
\end{equation}
 where $F_{i}(\vec{Q})$ and $F_{i}^{*}(\vec{Q})$ are the structure factor and conjugate structure factor of an $i$ block respectively. For brevity, let $F_{ij} = F_{i}(\vec{Q})F_{j}^{*}(\vec{Q})$ be the $ij$th components of what we will call the \emph{structure matrix} $\boldsymbol{F}$. Next, $G_{ij}(m_{3})$ denotes the probability that a $j$ block is $m_{3}$ blocks ahead of an $i$ block so we have that  $G_{ij}(m_{3}) = \pi_{i}{\xi}^{m_{3}}_{ij}$ for positive $m_{3}$. Postponing the case where $m_{3}$ is negative, we proceed by noting
 \begin{align}
     Y_{m_{3}}(\vec{Q}) &= \sum_{i \in B} \sum_{j \in B} F_{ij}\pi_{i}\xi_{ij}^{m_{3}}  \ , \ m_{3} > 0 \\
     &=
     \sum_{i \in B} \sum_{j \in B} F_{ji}^{T}\pi_{i}\xi_{ij}^{m_{3}} 
 \end{align}
and use that $\boldsymbol{F}$ is Hermitian, which is to say $\boldsymbol{F}^{T} = \boldsymbol{F}^{*}$, to deduce
\begin{align}
     Y_{m_{3}}(\vec{Q}) &=
     \sum_{i \in B} \sum_{j \in B} F_{ji}^{*}\pi_{i}\xi_{ij}^{m_{3}}
\end{align}
which we identify as
\begin{equation}
Y_{m_{3}}(\vec{Q}) = \Tr\bigg(\boldsymbol{F}^{*}\Diag(\pi)\boldsymbol{\Xi}^{m_{3}}\bigg)  
\end{equation}
where $\Tr$ is the trace.
Now, the probability of sampling an $i$ block from a crystal then finding a $j$ block $m_{3}$ blocks behind the first block, is equal to the probability of sampling a $j$ block from the crystal then finding a type $i$ block $m_{3}$ blocks ahead of the former block, consequently
\begin{equation}
Y_{-m_{3}}(\vec{Q}) = \Tr \bigg( \boldsymbol{F}\Diag(\pi)\boldsymbol{\Xi}^{m_{3}}\bigg)\ , \ m_{3} > 0.
\end{equation}
Now, continuing from equation \eqref{blehh}
\begin{align}
&\sum^{N_{c}}_{m_{3} = -N_{c}}(N_{c} - \abs{m_{3}})Y_{m_3}e^{2 \pi i m_{3} l} \label{SplitSum} \\
=&N_{c}\Tr\bigg(\boldsymbol{F}\Diag(\boldsymbol{\pi})\bigg) \nonumber \\ + &\sum^{N_{c}}_{m_{3} = 1} (N_{c} - m_{3})\Tr\bigg( \boldsymbol{F}^{*}\Diag(\boldsymbol{\pi})\boldsymbol{\Xi}^{m_{3}} \bigg)e^{2 \pi i m_{3} l} \nonumber \\
+ &\sum^{N_{c}}_{m_{3} = 1} (N_{c} - m_{3})\Tr\bigg( \boldsymbol{F}\Diag(\boldsymbol{\pi})\boldsymbol{\Xi}^{m_{3}} \bigg)e^{-2 \pi i m_{3} l}. \nonumber
\end{align}
Since the two summands on the RHS of the previous equation are complex conjugate, we seek an expression for only one of them, from which we can deduce the other easily. We do this by invoking the linearity of the trace to deduce
\begin{align}
&\sum^{N_{c}}_{m_{3} = 1} (N_{c} - m_{3})\Tr\bigg( \boldsymbol{F}^{*}\Diag(\boldsymbol{\pi})\boldsymbol{\Xi}^{m_{3}} \bigg)e^{2 \pi i m_{3} l} \label{Trace} \\
= &\Tr\bigg(  \boldsymbol{F}^{*}\Diag(\boldsymbol{\pi})\sum^{N_{c}}_{m_{3} = 1}(N_{c} - m_{3})\boldsymbol{\Xi}^{m_{3}}e^{2 \pi i m_{3} l} \bigg) \nonumber 
\end{align}
then we consider 2 cases, the first is for $\boldsymbol{\Xi}$ a diagonalisable matrix, where
\begin{align}
&\sum^{N_{c}}_{m_{3} = 1}(N_{c} - m_{3}) \boldsymbol{\Xi}^{m_{3}}e^{2 \pi i m_{3} l} \nonumber \\
=& \sum^{N_{c}}_{m_{3} = 1}(N_{c} - m_{3}) \boldsymbol{Q}
\left( \begin{array}{ccc}
\lambda_{1}^{m_{3}} & & 0  \\
 & \ddots &  \\
 0 & & \lambda_{n}^{m_{3}} \end{array} \right)
\boldsymbol{Q}^{-1}e^{2 \pi i m_{3} l} \\
=& \boldsymbol{Q} \\ &\sum^{N_{c}}_{m_{3} = 1}
\left( \begin{array}{ccc}
(N_{c} - m_{3})\lambda_{1}^{m_{3}}e^{2 \pi i m_{3} l} & & 0  \\
 & \ddots &  \\
 0 & & (N_{c} - m_{3})\lambda_{n}^{m_{3}}e^{2 \pi i m_{3} l} \end{array} \right) \nonumber \\
&\boldsymbol{Q}^{-1} \nonumber
\end{align}
for some invertible matrix $\boldsymbol{Q}$.
Now the sum
\begin{align}
s_{n} = \sum^{N_{c}}_{m_{3} = 1}(N_{c} - m_{3})\lambda_{n}^{m_{3}}e^{2 \pi i m_{3} l}
\end{align}
has analytic expression
\begin{align}
s_{n} = 
\begin{cases}
\frac{N_c}{2}(N_c - 1) \text{ if $\lambda_n e^{2 \pi i l} = 1$ }  \\
\frac{\lambda_n e^{2 \pi i l} ( \lambda_n^{N_c} e^{2 \pi i l N_c} + N_c(1 - \lambda_n e^{2 \pi i l} ) - 1)}{(1 - \lambda_n e^{2 \pi i l})^2} \text{ otherwise,}
\end{cases}
\end{align}
so we can let $\boldsymbol{\hat{S}}$ be the diagonal matrix with components $s_n$. 

Next, we consider the case of a defective $\boldsymbol{\Xi}$. Though we cannot diagonalise $\boldsymbol{\Xi}$, we can express it terms of its Jordan Canonical form
\begin{align}
\boldsymbol{J} = \boldsymbol{Q}^{-1} \boldsymbol{\Xi} \boldsymbol{Q}
\end{align}
where $\boldsymbol{J}$ is a block diagonal matrix comprised of Jordan blocks $\boldsymbol{J}_p$; which are themselves upper triangular matrices each associated with an eigenvalue $\lambda_p$ of $\boldsymbol{\Xi}$. The columns of the matrix $\boldsymbol{Q}$ are the eigenvectors and generalised eigenvectors of $\boldsymbol{\Xi}$, so $\boldsymbol{Q}$ is necessarily invertible. We proceed by first noting  
\begin{align}
&\sum^{N_{c}}_{m_{3} = 1}(N_{c} - m_{3}) \boldsymbol{\Xi}^{m_{3}}e^{2 \pi i m_{3} l} \nonumber \\
=& \sum^{N_{c}}_{m_{3} = 1}(N_{c} - m_{3}) \boldsymbol{Q}
\left( \begin{array}{ccc}
J_{1}^{m_{3}} & & 0  \\
 & \ddots &  \\
 0 & & J_{p}^{m_{3}} \end{array} \right)
\boldsymbol{Q}^{-1}e^{2 \pi i m_{3} l} \\
=& \boldsymbol{Q} \\ &\sum^{N_{c}}_{m_{3} = 1}
\left( \begin{array}{ccc}
(N_{c} - m_{3})J_{1}^{m_{3}}e^{2 \pi i m_{3} l} & & 0  \\
 & \ddots &  \\
 0 & & (N_{c} - m_{3})J_{p}^{m_{3}}e^{2 \pi i m_{3} l} \end{array} \right) \nonumber \\
&\boldsymbol{Q}^{-1}. \nonumber
\end{align}
Now the weighted sum of Jordan blocks
\begin{align}
\mathcal{S}_{p} = \sum^{N_{c}}_{m_{3} = 1}(N_{c} - m_{3})J_{p}^{m_{3}}e^{2 \pi i m_{3} l}
\end{align}
has expression
\begin{align}
\mathcal{S}_{p} &=  
\boldsymbol{J}_p e^{2 \pi i l} ( \boldsymbol{J}_p^{N_c} e^{2 \pi i l N_c} + N_c(\boldsymbol{I} - \boldsymbol{J}_p e^{2 \pi i l} ) - \boldsymbol{I}) \\
&\times (\boldsymbol{I} - \boldsymbol{J}_p e^{2 \pi i l})^{-2}
\end{align}
under the condition that the resolvent $\big( \boldsymbol{I} - \boldsymbol{J}_{p}e^{2\pi i l} \big)$ is invertible, which is true unless $\lambda_p e^{2 \pi i l} = 1$. In the case of an invertible resolvent, it is noteworthy for ease of computation that 
\begin{align}
    \bigg( \boldsymbol{I} - \boldsymbol{J}_p e^{2 \pi i l} \bigg)^{-1}_{ij}
    = 
    \begin{cases}
    (1 - \lambda_{p} e^{2 \pi i l })^{-1 + i - j} &\text{ if $i \leq j$} \\
    0 &\text{ otherwise }
    \end{cases}
\end{align}
and that 
\begin{align}
    \bigg(\boldsymbol{J}_p e^{2 \pi i l}\bigg)^{N_c}_{ij} = 
    \begin{cases}
    (\lambda_p e^{2 \pi i l})^{N_c} \lambda^{i - j} \binom{N_c}{j - i} &\text{ if $i \leq j$} \\
    0 &\text{ otherwise }
    \end{cases}
\end{align}
where $\binom{a}{b}$ are binomial coefficients. 
On the other hand, if $\lambda_p e^{2 \pi i l} = 1$ then the resolvent $\big( \boldsymbol{I} - \boldsymbol{J}_{p}e^{2\pi i l} \big)$ is singular and we note 
\begin{align}
    \bigg( \sum_{m_3 = 1}^{N_c}(N_c - m_3)J^{m_3}_p \bigg)_{ij} = \sum_{m_3 = 1}^{N_c}(N_c-m_3) \binom{m_3}{j-i}.
\end{align}
We can now define the matrix $\boldsymbol{\hat{S}}$ as the block diagonal matrix composed of the blocks $\mathcal{S}_p$

Having considered both cases of $\boldsymbol{\Xi}$ diagonalisable and not, we proceed by recalling equation \eqref{Trace} and note
\begin{align}
&\sum^{N_{c}}_{m_{3} = 1} (N_{c} - m_{3})\Tr\bigg( \boldsymbol{F}^{*}\Diag(\boldsymbol{\pi})\boldsymbol{\Xi}^{m_{3}} \bigg)e^{2 \pi i m_{3} l} \\
=& \Tr\bigg(\boldsymbol{F}^{*}\Diag(\boldsymbol{\pi})\boldsymbol{Q}^{-1}\boldsymbol{\hat{S}}\boldsymbol{Q}\bigg) \nonumber
\end{align}
hence we can express equation \eqref{SplitSum}
\begin{align}
&\sum_{m_{3} = -N_{c}}^{N_{c}}(N_{c} - \abs{m_{c}})Y_{m_{3}}e^{2 \pi i m_{3} l} \\
=&\Tr\bigg(\boldsymbol{F}\Diag(\boldsymbol{\pi}) \bigg) \nonumber \\
+&\Tr\bigg(\boldsymbol{F}^{*}\Diag(\boldsymbol{\pi})\boldsymbol{Q}^{-1}\boldsymbol{\hat{S}}\boldsymbol{Q}\bigg) \nonumber \\
+&\Tr\bigg(\boldsymbol{F}\Diag(\boldsymbol{\pi})\boldsymbol{Q}^{-1}\boldsymbol{\hat{S}}^{*}\boldsymbol{Q}\bigg) \nonumber
\end{align}
then using equation \eqref{BerlinerExpressionAppendix}, linearity of the trace, and the trace operator's cyclic permutation property
\begin{equation}
\Tr(\boldsymbol{ABC}) = \Tr(\boldsymbol{CBA}), 
\end{equation}
we deduce
\begin{align}
\frac{d\sigma}{d\Omega} =&\frac{\sin^2(N_a\pi h)}{\sin^2(\pi h)}\frac{\sin^2(N_b\pi k)}{\sin^2(\pi k)} \times \\
&\Tr\bigg(\Diag(\boldsymbol{\pi})(N_c\boldsymbol{F} + \boldsymbol{Q}^{-1}\boldsymbol{\hat{S}}\boldsymbol{Q}\boldsymbol{F}^{*} + \boldsymbol{Q}^{-1}\boldsymbol{\hat{S}}^{*}\boldsymbol{Q}\boldsymbol{F})\bigg). 
\nonumber
\end{align}
Finally we deploy the change of basis
\begin{align}
\Diag(\boldsymbol{\pi}) &= \boldsymbol{Q}^{-1}\boldsymbol{\hat{P}}\boldsymbol{Q} \\
\boldsymbol{F} &= \boldsymbol{Q}^{-1}\boldsymbol{\hat{F}}\boldsymbol{Q}
\end{align}
and once again use the linearity and cyclic permutation property of the trace to deduce
\begin{align}
\frac{d\sigma}{d\Omega} =\frac{\sin^2(N_a\pi h)}{\sin^2(\pi h)}\frac{\sin^2(N_b\pi k)}{\sin^2(\pi k)} \times \\
\Tr\bigg(\boldsymbol{\hat{P}}(N_c\boldsymbol{\hat{F}} + \boldsymbol{\hat{S}}\boldsymbol{\hat{F}}^{*} + \boldsymbol{\hat{S}}^{*}\boldsymbol{\hat{F}})\bigg)
\nonumber \\ =
\frac{\sin^2(N_a\pi h)}{\sin^2(\pi h)}\frac{\sin^2(N_b\pi k)}{\sin^2(\pi k)}
Re \bigg\{ \Tr \big(\boldsymbol{\hat{P}}(2\boldsymbol{\hat{S}}+N_c\boldsymbol{I})\boldsymbol{\hat{F}})\big) \bigg\}.
\end{align}

\section{Analytic derivatives of the scattering cross section} 
\label{dixB}
We may be interested in using scattering data to measure certain molecular quantities encoded in the structure factor. For example, the separation between a particular pair of atoms in a block of type $i$. We can represent these unknown molecular quantities as free parameters that we attempt to estimate by finding values for them that best fit experimental scattering data. This sort of refinement often requires the evaluation of a Jacobi matrix of derivatives. Consequently, it may be useful to have an expression for the analytic derivatives of the scattering cross section with respect to the structure factor's free parameters. Letting $\tau$ be such a free parameter, we have that

\begin{align}
\frac{\partial}{\partial \tau}\bigg(\frac{d\sigma}{d\Omega}\bigg) = &N_{a}N_{b} \delta(h - h_{0})\delta(k - k_{0}) \times \\
&\Tr\bigg(\Diag(\boldsymbol{\pi})(\partial \boldsymbol{F} + \boldsymbol{Q}^{-1}\boldsymbol{S}\boldsymbol{Q}\partial\boldsymbol{F}^{*} + \boldsymbol{Q}^{-1}\boldsymbol{S}^{*}\boldsymbol{Q}\partial \boldsymbol{F})\bigg)
\nonumber
\end{align}
where $\partial\boldsymbol{F}$ is a matrix with elements
\begin{equation}
\partial F_{ij} = \frac{\partial F_{i}}{\partial \tau} F^{*}_{j} + F_{i}\frac{\partial F^{*}_{j}}{\partial \tau}.
\end{equation}
Here, the derivative $\frac{\partial F_{i}}{\partial \tau}$ may be expressed analytically if possible or approximated using a finite difference if necessary.

\end{appendix}
\bibliographystyle{iucr}
\bibliography{References}

\end{document}